\documentclass[preprint,showpacs,preprintnumbers]{revtex4}
\usepackage{epsfig}
\usepackage{mathrsfs}
\usepackage{amsmath}
\usepackage{amssymb}
\usepackage{amsfonts}

\usepackage{color}
\definecolor{gxhighlight}{rgb}{0.85,0.9,1}

\newcommand{\bea}{\begin{eqnarray}}
\newcommand{\eea}{\end{eqnarray}}
\newcommand{\p}{\partial}
\newcommand{\LL}{\mathcal {L}}
\newcommand{\ov}{\over}
\newcommand{\dd}{\textrm{d}}

\begin{document}





\title{Comments on  Baryon  Melting in  Quark Gluon Plasma \\ with gluon condensation}
\author{Sang-Jin Sin}
\email{sangjin.sin@gmail.com} \affiliation{\vspace{0.3cm}
    Department of Physics, Hanyang University, Seoul 133-791, Korea
}
\author{Shuo Yang, Yang Zhou}
\email{yangshuo@itp.ac.cn, yzhou@itp.ac.cn} \affiliation{\vspace{0.3cm}
     Institute of Theoretical Physics, Chinese Academy of
    Sciences, Beijing 100190, P.R.China
}






\begin{abstract}
 We consider a black hole solution with a non-trivial dilaton from IIB super gravity
 which is expected to describe a  strongly coupled  hot gauge plasma with non-vanishing gluon condensation present.
 We   construct a rotating  and moving baryon to probe the screening and phases of the  plasma.
 Melting of the baryons in hot plasma in this background had been studied previously, however,  we show that  baryons melt much lower temperature than has been suggested previously.     \end{abstract}

\maketitle

\section{Introduction}
The gauge/string duality~\cite{AdS/CFT} has generated much interest in investigating quantum chromodynamics (QCD) at finite
temperature.
Interesting quantities of hot QCD plasma, such as
$\eta/s$, meson and baryon screening length ($L_s$), jet quenching parameter $\hat q$ have been obtained by gravity calculation, while they are hard  to compute in lattice.
The  data  from heavy ion collisions at RHIC shows that, above a critical temperature $T_c\sim170$MeV
(deconfinement transition), the QCD plasma is like a strongly coupled liquid~\cite{RHIC} with approximately ideal hydrodynamics. On the other hand, lattice results indicate that, the thermodynamics of QCD plasma is scale invariant
upto $\sim2T_c$~\cite{Karsch:2006xs}. Although these  points provide  plausibility for describing QCD plasma by a super Yang-Mill theory with a gravity dual with conformal invariance,
 to capture the behavior of real QCD,  one needs to extend the  gauge/gravity duality to nonconformal cases and there have been much efforts  ~\cite{Gubser:2008yx,Gursoy:2008bu} along this  direction.
  It is remarkable  that the screening length and jet quenching parameter can be affected by the nonconformality by $  20\sim30\%$  ~\cite{Liu:2008tz}.

One of the problem in holographic QCD is that
the temperature dependence  of the baryons are usually suppressed
in  holographic QCD, although it has been known that  heavy quark bound states can survive in quark gluon plasma~\cite{de Forcrand:2000jx} in real QCD.
In fact, for the most of the background
with regular horizon, the finite energy configuration of the
baryon vertex does not exist. However, in \cite{Ghoroku:2005tf}, it was found that in the system of non-extremal
$D3/D_{-1}$, closed baryon vertex operator exists even at finite temperature. Therefore this background is ideal to study the temperature dependence of various quantities around the critical temperature. Especially interesting one is the baryon   melting temperature and the associated phases.

Baryons in hot plasma in this background had been studied previously  ~\cite{Liu:2008bs,Li:2008py,Zhou:2008vf,Sin:2009dk,Ghoroku:2008tg}. In  \cite{Liu:2008bs,Li:2008py}, one baryon was considered as a
 point vertex with hanging strings.  It was shown that the velocity dependence of screening distance goes like  $L_s\sim (1-v^2)^{1/4}$.
 Baryons which have radius larger than $L_s$ will dissociate.
Since $L_s\sim 1/T$,  for any fixed baryon radius $R$,
 there is a critical temperature T$_c\sim 1/R$  above which baryons will dissociate.
In \cite{Zhou:2008vf}, the same analysis was made by taking into account the shape of baryon.
In \cite{Sin:2009dk}, we made screening related analysis using the non-conformal background found in \cite{Ghoroku:2005tf}.

Recently, the authors of \cite{Ghoroku:2008tg} suggested another melting temperature $T_m$ above  which no compact D5 vertex exists by studying the DBI action of D5 vertex.
In this paper,  we study thermodynamics of this model
in the presence of baryon vertex operators and show that
the baryons melt at much  lower temperature than  it was suggested in \cite{Ghoroku:2008tg} by comparing the free energy of baryon configuration and that of free string configuration.
 We also  examine the phases of a baryon in various  motions like rotating and  moving   configurations.

\section{ D3/D-instanton background and gluon condensation}

The gravity theory dual to the thermal four-dimensional gauge theory is a solution of 10D Type-IIB supergravity under
the Freund-Rubin ansatz for self-dual five form field strength~\cite{Ghoroku:2005tf,Kehagias:1999iy,Liu:1999fc}. In
string frame, the solution can be written as follow
\bea\label{metric}
\begin{split}
 e^{-{1\over 2}\phi}\textrm{d}s_{10}^2&=
 -{r^2\over r_+^2}\biggr(1-{r_0^4\over r^4}\biggr)\dd t^2+{r^2\over r_+^2}\dd x_i\dd x^i
 +{1\over 1-{r_0^4\over r^4}}{r_+^2\over r^2}\dd r^2+r_+^2\dd\Omega_5^2,\\
\end{split}\eea with a dilaton  and an axion \bea e^\phi=1+{q\over r_0^4}\log{1\over 1-{r_0^4\over r^4}},\quad
\chi=-e^{-\phi}+\chi_0,
\eea
 where $i=1,2,3$ and $q$ is gauge fields condensate parameter. $\phi$ and $\chi$ denote the dilaton and
the axion respectively. This metric includes an AdS black hole times a five-dimensional sphere, the dilaton and axion
depending on $r$. $r_+$ is the curvature radius of the AdS metric, $r$ is the coordinate of the fifth dimension of
AdS$_5$ and $r_0$ is the position of black hole horizon. The temperature of the gauge theory is given by Hawking
temperature of the black hole, $T={r_0\over \pi r_+^2}$. By duality, the gauge theory parameters $N_c$
and $\lambda$ ( t'Hooft coupling  ) are given by $
 \sqrt{\lambda}={r_+^2\over \alpha'}, {\lambda\over N_c}=g_{YM}^2=4\pi g_s
$, where ${1\over 2\pi\alpha'}$ is string tension and $g_s$ is the string coupling constant. The self-dual
Ramond-Ramond field strength is \bea\begin{split}
 F_{(5)}=\dd C_{(4)}=4r_+^4\Omega_5\dd\theta_1\wedge...\wedge \dd\theta_5-4{r^3\over r_+^4}\dd t\wedge \dd x_1\wedge \dd x_2\wedge \dd x_3\wedge \dd r,
\end{split}\eea where $\Omega_5=\sin^4\theta_1\sin^3\theta_2\sin^2\theta_3\sin\theta_4$.

The baryon construction in gravity involves $N_c$ fundamental strings with the same orientation, beginning at the heavy
quarks on the flavor brane and ending on the baryon vertex in the interior of bulk geometry~\cite{Witten:1998}, which
is a D5 brane wrapped on the S$^5$ ( AdS$_5\times$S$^5$ background ). \footnote{Generally, $N_c$ quarks are allowed to
be placed at arbitrary positions in $\vec x$ space on the boundary. Note that these quarks are heavier than component
quarks of light mesons, and the $N_c$ quarks bound states can not easily be considered as an effective field on the
boundary ( fluctuations of flavor brane in the picture with flavor )~\cite{Li:2008py}.} The D5 brane carries a radial
U(1) flux and wraps the S$^5$ with radial extension. The action of D5 brane includes DBI action plus Chern-Simons
action, given by \bea\begin{split}
 S_{D5}=&-T_5\int \dd^6\sigma e^{-\phi}\sqrt{-\det(g_{ab}+2\pi \alpha'F_{ab})}\\
 &+T_52\pi\alpha'\int A_{(1)}\wedge
 \mathscr{P}(F_{(5)}),
\end{split}\eea where the 6D world volume induced metric $g_{ab}=\p_aX^\mu\p_bX^\nu G_{\mu\nu}$, and the pull back of five form
$\mathscr{P}(F_{(5)})=\p_{a_1}X^{\mu_1}...\p_{a_4}X^{\mu_5}F_{\mu_1...\mu_5}$. The D5 brane tension
 $T_5={1\over g_s(2\pi)^5l_s^6}$, and the world volume field strength of U(1) flux $F_{(2)}=\dd A_{(1)}$.
The Chern-Simons term endows D5 brane with U(1) charge. By the following consistent ansatz that describes the embedding
D5 brane \bea\begin{split} \tau=t,\quad\sigma_1&=\theta,\quad \sigma_2=\theta_2,...\sigma_5=\theta_5,\\
 r&=r(\theta),\quad x=x(\theta),
 \end{split}\eea 
 we see only SO(5) symmetric configurations of D5 brane which stand for baryons in 4D real spacetime ($t,\vec x$) are considered,
 and the embedding function can be determined by $r(\theta)$ and $x(\theta)$. The gauge field on D5 can also be written as
 $A_t(\theta)$ for symmetry. The action of D5 brane is given by
 \bea\begin{split}
 S=T_5\Omega_4r_+^4\int \dd t \dd\theta \sin^4\theta \biggr[4A_t-e^{\phi\over 2}\times
 \sqrt{(1-r_0^4r^{-4})r^2+r'^2+
 (1-r_0^4r^{-4})r^4 r_+^{-4}x'^2-F_{\theta t}^2}\biggr]
 \,,
 \end{split}
 \eea where $\Omega_4=8\pi^2/3$ is the volume of unit four sphere. To obtain the configuration of D5 brane, we should
 solve the gauge field at first. The equation of motion turns to be
 \bea
 \p_{\theta} D=-4\sin^4\theta.
 \eea The solution to the above equation is
 \bea
 \begin{split}
 D(\nu,\theta)={3\over 2}(\sin&\theta\cos\theta-\theta+\nu\pi)+\sin^3\theta\cos\theta,\\
 &0\leq\nu={k\over N_c}\leq1,
 \end{split}\eea where $k$ denotes the number of Born-Infeld strings emerging from south pole of S$^5$. More details
 about this explanation can be found in~\cite{CGST}. To eliminate the gauge field in favor of $D$, we shall transform
 the original Lagrangian to obtain an energy functional of the embedding function as follow
 \bea\label{rxL}
 \begin{split}
 \mathcal {H}
 =\tilde{T}\int \dd\theta e^{\phi\over 2}\sqrt{\biggr(1-{r_0^4\over r^4}\biggr)r^2+r'^2+
 \biggr(1-{r_0^4\over r^4}\biggr){r^4\over r_+^4}x'^2}
 \times\sqrt{D^2+\sin^8\theta}\;.
\end{split}
 \eea where $\tilde{T}=T_5\Omega_4r_+^4$.
In order to find the configuration of D5 brane, one must extremize $\mathcal {H}$, with respect to $r(\theta)$ and
$x(\theta)$ respectively. A closed solution of D5 is  identified to be a physical baryon vertex. More discussion about these
solutions can be found in~\cite{GI0801}.
 By solving equation of motion for $r(\theta)$ and $x(\theta)$ in (\ref{rxL}), we can find different kinds of
solutions for baryon vertex.

\subsection{Baryon vertex solutions} Note that point vertexes in real spacetime
corresponds to $x'(\theta)=0$.
 If $q=0$, the gravity
theory is the usual AdS black hole. In that case, vertex D5 brane with a DBI+CS action can not have a closed solution.
Here, choosing $q>0$(or some critical value) to keep D5 solutions closed.
 The solutions
are independent on $r_+$, if $x'(\theta)=0$. Only two parameters $q$ and $r_0$ determine behaviors of solutions. When
we choose suitable parameters $q=2$ and, $r_0$ between 0.1 and 0.689, the vertex brane solutions can have four typical
behaviors  according to the initial value $r(0)$~\cite{GI0801}.  They correspond to four types  of configurations  of a baryon. From these solutions, we see that there is always a singularity in
$r_e=r(\pi)$, if we give
 initial conditions $ r'(0)=0,r(0)=C$,
where $C$ is a constant.

\subsection{Force balance condition} Adding fundamental
strings can help to eliminate this singularity and keep charge conserved. For simplicity, consider that $N_c$
fundamental strings all attach the north pole of S$^5$, which means $\nu=0$. $N_c$ static quarks are arranged on a
circle in ($x_1,x_2$) space, whose coordinates can also be written as ($\rho,\alpha$). By the following consistent
ansatz that describes the embedding fundamental strings $
 \tau=t,\quad \sigma=r,\quad \rho=\rho(r),
$  we write the string action $
 S_F
 ={1\over 2\pi\alpha'}\int \dd t \dd r \LL_F
 $. To
eliminate the singularity of cusp of D5 brane at $r_e$, one needs force balance conditions. One force balance condition
in $\rho$ direction is satisfied for central symmetry. Another force balance condition in $r$ direction is given by
\bea \label{FBC}
  N_c\biggr\{\LL_F -  \rho' {\p \LL_F \over \p \rho'}\biggr\}\biggr|_{r_e} = 2\pi \alpha'{\p \mathcal {H} \over \p
   r_e}\;. \eea The left hand of equation (\ref{FBC}) is up force of string and the right hand is down force of
   brane. The balance point is the singularity of vertex solution.

\section{ Phases of  baryon in various motion}
\subsection{Baryon in motion and the Binding energy}
A baryon is consist of compact D5 brane and $N_c$ strings coming out of the north pole of it.
The latter is considered as quarks and we shall consider such quarks moving in medium and rotating in a plane, corresponding to boosted and high spin hadron state respectively \footnote{Actually, the general shape of multi quarks is a sphere in 3D space for largest symmetry, rather than circular.}.
The medium wind will effect the vertex brane and fundamental strings in the same time, and the vertex brane can not feel the rotating effect, because it is a central point in the rotation plane. We consider quarks moving in $x_3$ direction and rotating in ($\rho,\alpha$) plane. For simplicity, we shall stand in the rest frame of the baryon configuration. The metric (\ref{metric}) can be boosted such that it describes a gauge plasma moving with a wind velocity $v$ in the negative $x_3$-direction. The boosted metric is given by \bea\begin{split}
 e^{-{1\over 2}\phi}ds_{10}^2=-A\dd t^2+2B\dd t\dd x_3+C\dd x_3^2
 +{r^2\over r_+^2}(\dd\rho^2+\rho^2\dd\alpha^2)+{r_+^2\over r^2}{1\over f(r)}\dd r^2+r_+^2\dd\Omega_5^2
\end{split}\eea where \bea
 A={r^2\over r_+^2}\biggr(1-{r_1^4\over r^4}\biggr),B={r_1^2r_2^2\over r^2r_+^2},C={r^2\over r_+^2}\biggr(1+{r_2^4\over r^4}\biggr)
\eea with \bea\begin{split}
 r_1^4&=r_0^4\cosh^2\eta, \quad r_2^4=r_0^4\sinh^2\eta,\\
  v&=-\tanh\eta,\quad f=1-{r_0^4\over r^4}.
\end{split}\eea In the boosted metric, baryon configuration will depend on $\eta$. Both vertex brane and fundamental string
solutions will be different from the original ones with $\eta=0$. First, we pay attention to vertex brane solutions.
For point brane vertex, one notes that $x^i$ is independent on $\theta$. The D5 brane action in boosted metric is given
by \bea\begin{split} \mathcal {H}_\eta=\tilde{T}\int \dd\theta e^{\phi\over
2}\sqrt{\biggr(r^2+{1\over f}r'^2\biggr)\biggr(1-{r_0^4\cosh^2\eta\over r^4}\biggr)}
\times\sqrt{D^2+\sin^8\theta}\;.
\end{split}\eea To obtain rotating fundamental string configuration, we give the following consistent ansatz of embedding
function \bea
 \tau=t,\quad\sigma=r,\quad\alpha=\omega t,\quad\rho=\rho(r).
\eea Facing the wind in $x_3$ direction, quarks arranged on the circle in $x_1-x_2$ plane will keep staying in
$x_1-x_2$ plane and stand on a circle, because they all have the same force.  Then the rotating string action in the
boosted metric can be written as \bea\tilde{S}_F={\mathcal {T}\over 2\pi\alpha'}\int_{r_e}^{r_\Lambda}\dd
r\tilde{\LL}_F\;,\eea where the Lagrangian \bea \tilde{\LL}_F=e^{\phi\over 2}\sqrt{\biggr(1-{r_0^4\cosh^2\eta\over
r^4}-\rho^2\omega^2\biggr)\biggr({1\over f}+{r^4\over r_+^4}\rho'^2\biggr)}. \eea
 To solve the equation of motion of strings, we need two initial conditions. One is known by
$\rho(r_e)=0$ for symmetry, and the other must be calculated by the force balance condition (\ref{FBC}). To get the
baryon radius in the boundary, we define $
 L_q=\int_{r_e}^{r_\Lambda}\rho'(r)\dd r.
$ For $\eta>0,\omega=0$, string Lagrangian contains no $\rho$ and one can solve $\rho'$ from equation of motion of
$\rho$ and express baryon radius $L_q$ in terms of $r_e$ and $\eta$ by \bea
 L_q=\int_{r_e}^{r_\Lambda}\dd r {K(r_e,\eta)r_+^4(r^4-r_0^4)^{-1/2}\over \sqrt{e^\phi(r^4-r_0^4\cosh^2\eta)-K^2(r_e,\eta)r_+^4}}
\eea where $K(r_e,\eta)$ is constant coming from the equation of motion of $\rho(r)$, $\p_{\rho'}\tilde{\LL_F}=K$,
determined by the force balance condition \bea K(r_e,\eta)={\sqrt{(r_e^4-r_0^4\cos^2\eta)e^{\phi(r_e)}}\over
r_+^2\sqrt{1+ \theta'^{-2}r_e^{-2}f^{-1}}}\;. \eea For $\eta>0, \omega>0$,  equation of motion of $\rho$ is difficult to
solve analytically. One must search for the numerical result. Screening length is defined as the maximum of $L_q$ while
we change $r_e$.

 The total baryon energy is given by sum
 of the energy of $N_c$ strings
and that of the baryon vertex :  \bea\label{Etotal}
 E_{total}=N_cE_{string}+E_{D5},
\eea where the masses of string and vertex brane are given by\bea\label{stringenergy} E_{string}=\omega {\p \tilde{L}
\ov \p\omega}-\tilde{L}\;, \quad E_{D5}=\mathcal {H}_\eta\;,
 \eea
 Where $\tilde{L}=\frac {1}{2\pi \alpha'}\int_{r_e}^{r_{\Lambda}}\dd r \tilde{\LL}_F$ is the string Lagrangian.
We define the effective baryon binding  energy as the difference between the  total energy of  a baryon and that of $N_c$ quarks   rooted at the black hole horizon.
Notice that compact D5 brane  wrapping the horizon  has no mass.
If the radius
 of horizon is $r_0$,  and boundary is  at $r=r_{\Lambda} $,
 the mass of deconfined quark is given by
 \bea
  E_q={1\ov 2\pi\alpha'}\int_{r_0}^{r_\Lambda}e^{\phi/2}\dd r\;
 \eea
 Then the binding energy is given  by
  \bea
  E_I=E_{total}-N_cE_q\;.
 \eea
Concretely, $E_I$ is written by \bea\label{potential}\begin{split}
 E_I=&{N_c\over 2\pi\alpha'}\int_{r_e}^{r_\Lambda}\dd r {e^\phi\over \tilde{\LL}_F}\biggr({1\over f}+{r^4\rho'^2\over r_+^4}\biggr)\biggr(1-{r_0^4\cosh^2\eta\over
 r^4}\biggr)\\
 &+\mathcal {H}_\eta-{N_c\over 2\pi\alpha'}\int_{r_0}^{r_\Lambda}e^{\phi/2}\dd r\;.
\end{split}\eea

 \begin{figure}[tbp]
\centering
  \includegraphics[bb=8 567 582 778, width=12 cm, clip]{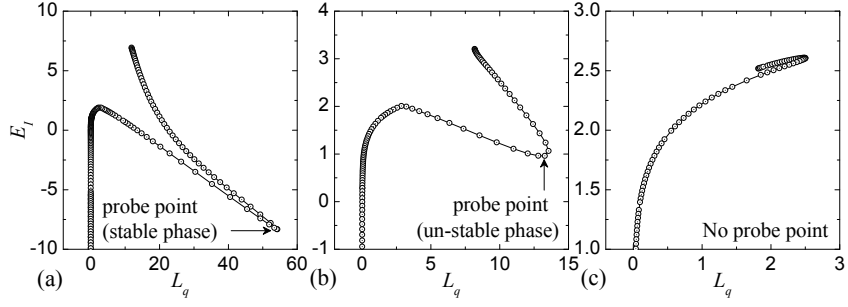}
  \caption{\small a) Potential curves at low temperature. b) Potential curves at middle temperature. c) Potential curves at high temperature.
  }\label{EI_lq}
\end{figure}

 \begin{figure}[tbp]
\centering
  \includegraphics[bb=27 601 374 782, width=10 cm, clip]{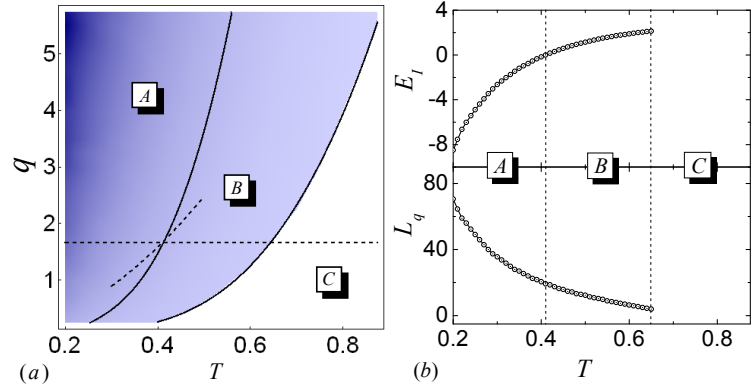}
  \caption{\small LEFT:  Stability of baryon:    A: Region of  stable baryon ($E_I<0$).
  B:  Un-stable baryon  ($E_I>0$).   C: Region where no baryon configuration is allowed.
  The horizontal dash line is $q=q_c$ and the short dash line is used to determine $q_c=1.67$.
  RIGHT: $T$ dependence of  bind energy and the baryon  size, with $q=q_c$.
  }\label{EI}
\end{figure}
 \begin{figure}[tbp]
\centering
  \includegraphics[bb=18 563 572 784, width=12 cm, clip]{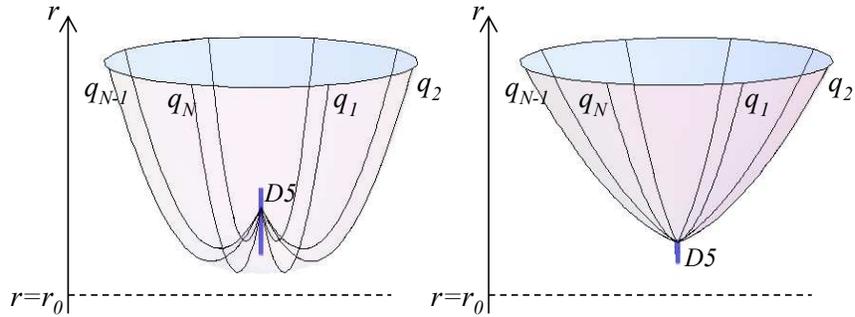}
  \caption{\small LEFT: Baryon configuration at low temperature, which corresponds to the case there is a minimal potential. RIGHT: Baryon configuration at high temperature, which exhibits monotonic potential.
  }\label{conf}
\end{figure}

\subsection{ Phases  of baryon}

Our gravity background is an AdS black hole modified by a dilaton and an axion. In the vacuum $T=0$, the dilaton
controls the effective coupling constant of QCD, and the axion is dual to the QCD $\theta$ angel~\cite{Gursoy:2008bu}.
By the calculation of Wilson loop, one can find that, the effective potential exhibits confining nature within a
temperature dependent screening length while for large separation of quark and anti quark  experience confining potential.  If we introduce a hard wall in IR regime by hand, we could introduce (gluon) confining background
IR~\cite{Ghoroku:2005tf}. However, such confinement transition by geometry changing is relevant to the
gluon dynamics and it is not directly related to  that of quarks and baryons.
Since our interest is the dynamics of the quarks/baryons, we do not introduce
a hard wall and therefore we do not have Hawking Page transition.

\subsubsection{Heavy probe baryon configuration and Critical temperature}
We shall focus on the effective potential between quarks of baryon, which was derived in (\ref{potential}). To make this part clear, we need some background knowledge. As is well known, in pure gauge theory, to understand confinement, we are particularly interested in the effective potential of gauge interaction between external fundamental probes. In thermal gauge theory and ones with flavors, the effective potential comes from more complex sources, where we should take account of finite temperature fluctuations and some flavor dynamics. Here we study the finite temperature case but leave the flavor dynamics to the future work. In particular, we focus on lower temperature backgrounds than the usual backgrounds where the potential curves ($E_I$ vs $L_q$) exhibit monotonic behaviors ( we only take the lower branch of potential curves ), which are shown in (c) in Fig.$\ref{EI_lq}$. In thermal plasma, we can think temperature $T$ is a typical scale. Now we focus on baryons with radius $L_q$ larger than a cutoff $L_c$, where $L_c\sim {1\over T}$. Note that the potential curve reflects the interaction of quarks of baryon at different separations. As $T$ decreases, we obtain the potential curves typically like (b) in Fig.$\ref{EI_lq}$. We consider that baryons in the bottom of the curves are more stable. Ignoring the infinitely negative piece of the left part of curves in (b) in Fig.$\ref{EI_lq}$ since the cutoff, we observe that there appears a minimal point of potential for each curve. In the following sections, we focus on baryon configurations at this point. We call this point ``probe point".
Since $E_I$ can also be defined as binding energy of baryon, for negative $E_I$, baryon is more stable than $N_c$ free quarks,  therefore
quarks  will bound into hadrons. Thus, we consider the ``probe point" in (a) in Fig.$\ref{EI_lq}$ is a really stable baryon. And critical temperature $T_c$ can be defined by ``probe point" with $E_I=0$.
\footnote{
We set $q=q_c$ to describe our background in the middle temperature region.
We compare our  background with that in work~\cite{Kajantie:2006hv}, and determine $q_c=1.67$ in Fig.\ref{EI}. }

 \begin{figure}[tbp]
\centering
  \includegraphics[bb=40 505 546 770, width=10 cm, clip]{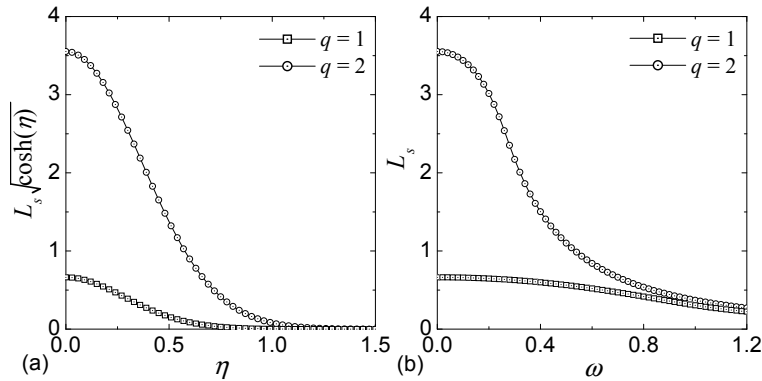}
  \caption{\small LEFT: velocity dependence of screening length.
  RIGHT: Rotation dependence of screening length.
  }\label{ls}
\end{figure}

In Fig.\ref{EI_lq}, we plotted potential curves  VS  baryon size ($E_I$ vs $L_q$). Notice  that there exist three typical potential curves
for low temperature (a), middle temperature (b) and high temperature (c).
Above $T_m(T_m>T_c)$, there is no stable baryon and no ``probe point".
So previously $T_m$ was considered as the melting temperature of baryons.
However, here we see that if temperature is higher than the $T_c$,  quark phase is the preferred one and therefore
baryon should melt before we reach the $T_m$.
We plotted the size and effective potential of ``probe points" states covering a wide temperature and condensation
region in Fig.\ref{EI} (LEFT).
We also plotted the baryon potential and size depending on $T$ in Fig.\ref{EI} (RIGHT). The reader may be puzzled about where the ``probe points" come from. The physical reason is that at low temperature, the projection line of vertex brane solution is not monotone increasing in $r$ direction as $\theta$ increases from $0$ to $\pi$. The configuration of the whole baryon is described by left one in Fig.$\ref{conf}$, which can be compared with the right normal one at higher temperature. Note that, in order to calculate the quark separation and effective potential of baryon for the left one in Fig.$\ref{conf}$, one should integrate two parts of one whole fundamental string, that is $r_e\rightarrow r_{min}$ and $r_{min}\rightarrow r_{max}$. 
\subsubsection{Screening effect at high temperature}
In higher temperature background, we have no stable baryon by potential study. However, we can obtain the screening effect, which provides the screening length of baryon (even unstable). According to the screening analysis in section III.A, for the boosted heavy baryon and high spin heavy baryon, we calculate the $\eta$ and
$\omega$ dependence of the screening length at $r_0$=0.7. As shown in Fig.\ref{ls}~\cite{Sin:2009dk}, gluon condensation makes the screening length larger than before. Baryon screening length obeys $L_s\sim(1-v^2)^{1/4}$ at large velocity.

\section{Conclusion}
In this letter, we probe the hot nonconformal QCD plasma by a heavy baryon, and find that
the baryons actually melt before it reach the ``melting temperature
$T_m$''
beyond which baryon configuration does not exist.
We determine the critical temperature $T_c$,
the quark state is energetically more favorable than the baryon state.  $T_c$  is lower than $T_m$.

We also calculated  the temperature dependence of the
free energies for each phase and the temperature dependence of baryon size. In the high temperature region, we calculate boost velocity $\eta$ and spin  dependence of screening length.


There are different approaches to the baryon in medium advocated in
\cite{yseo}. The idea is that flavor probe brane is deformed by attached strings which are emanating from the baryon vertex.
Since the fundamental string is more expensive than the deformed
probe brane or baryon vertex,
two branes are in contact with force balancing condition.
It would be very interesting to study the phase transition point $T_c$ in this picture.
Study in this direction is currently under progress.

One can also study the phase transition by considering the fermionic charges \cite{density} more explicitly.

\section{Acknowledgements}
The work of SJS was supported in part by KOSEF Grant R01-2007-
000-10214-0 and SRC Program of the KOSEF through the CQUeST with grant number
R11-2005-021.



\end{document}